\documentclass[referee]{raa}           

\usepackage{graphicx,times}
\usepackage{natbib}
\usepackage{amssymb,amsmath}
\bibpunct{(}{)}{;}{a}{}{,}

\usepackage[pagebackref=true]{hyperref}

\begin{document}

   \title{Study of 26 Galactic open clusters with extended main-sequence turnoffs}

 \volnopage{ {\bf 20XX} Vol.\ {\bf X} No. {\bf XX}, 000--000}
   \setcounter{page}{1}

   \author{Yang-Yang Deng\inst{1,2,3} and Zhong-Mu Li\inst{3}
   }

   \institute{ Yunnan Observatories, Chinese Academy of Sciences, Kunming 650216, China {\it dengyangyang@ynao.ac.cn}\\
        \and
             University of Chinese Academy of Sciences, Beijing 100049, China\\
	\and
Institute of Astronomy and Information, Dali University, Dali 671003, China\\
\vs \no
   {\small Received 20XX Month Day; accepted 20XX Month Day}
}

\abstract{Recent studies indicate that some Galactic open clusters (OCs) exhibit extended main-sequence turnoff (eMSTO) in their colour-magnitude diagrams (CMDs). However, the number of Galactic OCs with eMSTO structures detected so far is limited, and the reasons for their formation are still unclear. This work identifies 26 Galactic OCs with undiscovered eMSTOs and investigates the causes of these features. Stellar population types and fundamental parameters of cluster samples are acquired using CMD fitting methods. Among them, the results of 11 OCs are reliable as the observed CMDs are well-reproduced. 
We propose the crucial role of stellar binarity and confirm the importance of stellar rotation in reproducing eMSTO morphologies. The results also show that the impact of age spread is important, as it can adequately explain the structure of young OCs and fit the observed CMDs of intermediate-age OCs better. 
}

   \authorrunning{Y.-Y. D $\&$ Z.-M. L}            
   \titlerunning{Galactic OCs with eMSTOs}  
   \maketitle

%
\section{Introduction}           
\label{sect:intro}

Extended main-sequence turnoff (eMSTO) is a well-known phenomenon observed in the colour-magnitude diagram (CMD) of star clusters, inparticular for the globular clusters found in Magellanic Clouds (MCs). Recent works have shown that many Galactic clusters also exhibited eMSTO features in CMDs. The reasons for this structure have been studied in detail. For example, \cite{2018ApJ...863L..33M} first discovered the eMSTO structure in Galactic open cluster (OC) M11 and explained it as a result of stellar rotation. Other open clusters, i.e., NGC 2099, NGC 2360, and NGC 2818, also exhibited the eMSTO features. \cite{2018ApJ...869..139C} observed 12 Galactic OCs with eMSTOs and examined the effect of stellar rotation and internal age variation, concluding that stellar rotation was mainly responsible for the particular structure. Meanwhile, a series of work \citep[e.g.,][]{2019ApJ...876..113S, 2020MNRAS.491.2129D, 2021A&A...656A.149A, 2021AJ....162...64M} demonstrated that eMSTOs in Galactic OCs could mainly be attributed to the different rotation rates of the member stars. The faster rotators tended to be redder in colour and brighter in magnitude around the MSTO region.

Besides rotation, the eMSTO structure of Galactic OCs could also be explained by other alternative reasons, similar to the clusters in MCs. For instance, \cite{2019ApJ...887..199G} combined stellar rotation and age spread to clarify the existence of eMSTO in Milky Way OC NGC 2818 and three other star clusters in Large MC. While stars with rotation could create a similar appearance to the overall morphology of CMDs, the impact of extended star formation was necessary. The eMSTO of Milky Way OC Collinder 347 was corresponding to multiple stellar populations (SPs) with different ages \citep{2019MNRAS.490.2414P}. In very young clusters ($\sim$10-20 Myr), binary interaction played an important role in reproducing the eMSTO feature by combining the effects of variable stars and stellar rotation \citep{2019ApJ...876...65L}. A binary-driven formation mechanism \citep{2021MNRAS.502.4350S} was proposed for intermediate-age clusters (i.e., NGC 3960, NGC 6134, IC 4756, NGC 5822, and NGC 2818) as the main reason for eMSTO features, because the detection of the correlation between number fraction of slow rotators versus the total number of MSTO stars and their binary fraction. The impact of age spread, binary stars, and rotating stars on three clusters - M11, NGC 6819, and LP 585 were also explored in \cite{2022MNRAS.512.3992C}. It suggested that the eMSTO phenomenon observed in younger cluster ($\sim$0.3\,Gyr) was caused by the factors above. In comparison, rotation and binaries might contribute to the same phenomena in older clusters ($\sim$1.1 and 2.0\,Gyr).
In addition, the eMSTO feature of M37 was interpreted as the consequence of stellar rotation and chemical spread \citep{2022MNRAS.516.3631G, 2023MNRAS.524..108G},  possibly due to variations of [Fe/H] or helium. Consequently, there are multiple potential causes for the eMSTOs of Galactic OCs.

In recent years, hybrid machine learning and CMD fitting methods have been combined to identify over 10,000 Galactic star clusters \citep[][and references therein]{ 2023A&A...673A.114H, 2023MNRAS.526.4107P}.
The newly acquired view of the Milky Way provides clear CMDs for both known and new clusters, especially those with eMSTO morphologies. In order to investigate Galactic SPs further, we need to make clear the main causes of this feature. Our current work \citep{li2024bsec} identified 5411 Galactic cluster candidates, many of which exhibited eMSTO features and could be considered as excellent OC samples.

In this work, we present 26 Galactic OCs with eMSTOs and investigate the primary causes of these feature using CMD fitting methods. The paper is structured as follows. In Section 2, we briefly introduce the data source of the OC samples and summarize the characteristics of the data. Then, we present the CMD fitting method and process in Section 3. The fitting results and the main causes of eMSTOs are discussed in Section 4, while a summary is presented in Section 5.

\section{Cluster Samples Selection}
\label{sect:Obs}

We select cluster samples from our present work \citep{li2024bsec}, which utilizes the astrometric and photometric data from Gaia DR3 \citep{2023A&A...674A...1G}. 
This work puts forward a Blind Search-Extra Constraint (BSEC) method to hunt and identify the Galactic OCs. Apart from the use of both HDBSCAN \citep{10.1007/978-3-642-37456-2_14} and GMM \citep{article} methods to select cluster member stars with a probability greater than 0.9, the main feature of this work is the addition of an extra constraint. The member stars of detected OCs are redetermined using colour excess constraint, by eliminating those deviated from the specific curve in the ($G_{BP}$ - $G_{RP}$) versus ($G_{BP}$ - $G$) plane (colour-colour relation). The colour excess constraint is effective for constraining member stars when differential reddening in star cluster is less than 0.5\,mag. For more information, please refer to the literature. We extract 26 Galactic OCs that exhibit eMSTO phenomena in CMDs from the catalog and list the basic parameters in Table 1. Twenty clusters are newly discovered, while the known six ones have not reported eMSTO structures before. We check the CMDs of these clusters, the colour spreads around the TOs are narrower than 0.5\,mag. Thus the colour excess constraint is useful for identifying the member stars. 
We also revisit the distribution of proper motions of member stars in each cluster to examine if there are field stars that follow the colour-colour relation but with different distribution in the proper motion phase space. The results show that none of the proper motion distributions of the 26 clusters exhibits clearly separated bimodal distribution. Therefore, the CMDs of selected clusters in this work are cleaner than in those without extra constraint. The data for CMD fitting are magnitudes in $G$, $G_{BP}$, and $G_{RP}$ bands.

\begin{table}
	\caption{Basic parameters of the 26 clusters from \cite{li2024bsec}. $N_{*}$ is the number of member stars.}       
	\label{table:1}      
	\centering                          
	\begin{tabular}{c c c c c c c c}        
		\hline                
		ID & ra & dec &  $\varpi$ &$\mu_{\alpha}$~cos$\delta$ & $\mu_{\delta}$ & $N_{*}$ &  radius\\
		   & [deg]&[deg]&[mas]& [mas/yr]&[mas/yr]& &[deg]\\
		\hline 
		LSC0667  &  157.979$\pm$0.107  &  -57.021$\pm$0.068  &  0.541$\pm$0.038  &  -9.716$\pm$0.047  &  4.421$\pm$0.052  &71  &  0.233  \\
		LSC3348  &  141.845$\pm$0.146  &  -56.987$\pm$0.112  &  0.733$\pm$0.014  &  -7.752$\pm$0.059  &  5.759$\pm$0.060  &91  &  0.288  \\
		LSC3491  &  115.646$\pm$0.157  &  -16.340$\pm$0.127  &  0.105$\pm$0.047  &  -1.468$\pm$0.258  &  1.575$\pm$0.275  &322  &  0.345  \\
		LSC3803  &  103.307$\pm$0.169  &  10.693$\pm$0.174  &  0.248$\pm$0.127  &  -0.589$\pm$0.108  &  -0.860$\pm$0.278  &356  &  0.417  \\
		LSC4611  &  124.340$\pm$0.166  &  -30.035$\pm$0.174  &  0.222$\pm$0.113  &  -2.461$\pm$0.271  &  2.988$\pm$0.105  &493  &  0.387  \\
		LSC5048  &  219.652$\pm$0.081  &  -62.151$\pm$0.032  &  0.313$\pm$0.059  &  -3.450$\pm$0.121  &  -1.712$\pm$0.091  &87  &  0.111  \\
		LSC5120  &  224.303$\pm$0.136  &  -62.573$\pm$0.094  &  0.370$\pm$0.052  &  -4.424$\pm$0.128  &  -4.249$\pm$0.129  &450  &0.346  \\
		LSC5210  &  191.036$\pm$0.130  &  -58.098$\pm$0.055  &  0.547$\pm$0.067  &  -2.204$\pm$0.107  &  -1.220$\pm$0.100  &154  &0.206  \\
		LSC5301  &  108.200$\pm$0.173  &  -4.906$\pm$0.161  &  0.139$\pm$0.067  &  -0.890$\pm$0.148  &  0.402$\pm$0.267  &354  &0.405  \\
		LSC5559  &  131.156$\pm$0.088  &  -35.895$\pm$0.081  &  0.333$\pm$0.024  &  -2.608$\pm$0.040  &  5.666$\pm$0.044  &135  &0.249  \\
		LSC5867  &  77.723$\pm$0.164  &  40.047$\pm$0.159  &  0.233$\pm$0.125  &  0.478$\pm$0.266  &  -0.187$\pm$0.142  &208  &0.394  \\
		LSC0398  &  107.769$\pm$0.176  &  -6.476$\pm$0.168  &  0.250$\pm$0.127  &  -0.865$\pm$0.096  &  0.587$\pm$0.288  &455  &0.427  \\
		LSC0574  &  119.172$\pm$0.177  &  -25.125$\pm$0.165  &  0.224$\pm$0.128  &  -1.748$\pm$0.271  &  2.439$\pm$0.103  &605  &0.399  \\
		LSC1134  &  120.966$\pm$0.170  &  -20.028$\pm$0.178  &  0.196$\pm$0.113  &  -1.386$\pm$0.254  &  1.098$\pm$0.121  & 283  &0.391  \\
		LSC1323  &  113.097$\pm$0.153  &  -8.869$\pm$0.147  &  0.198$\pm$0.107  &  -0.871$\pm$0.095  &  0.808$\pm$0.275  & 294  &0.431  \\
		LSC4612  &  129.249$\pm$0.175  &  -29.224$\pm$0.140  &  0.242$\pm$0.114  &  -2.648$\pm$0.109  &  2.856$\pm$0.284  & 181  &0.388  \\
		LSC5576  &  112.340$\pm$0.175  &  -10.282$\pm$0.177  &  0.231$\pm$0.125  &  -1.148$\pm$0.248  &  0.838$\pm$0.109  & 470  &0.403  \\
		LSC0607  &  106.496$\pm$0.175  &  -28.987$\pm$0.171  &  0.176$\pm$0.112  &  -0.795$\pm$0.103  &  1.938$\pm$0.259  & 399  &0.383  \\
		LSC0623  &  118.493$\pm$0.176  &  -44.822$\pm$0.176  &  0.084$\pm$0.041  &  -1.348$\pm$0.245  &  2.870$\pm$0.276  & 576  &0.360  \\
		LSC2064  &  104.111$\pm$0.151  &  -30.365$\pm$0.148  &  0.221$\pm$0.125  &  -0.344$\pm$0.257  &  2.593$\pm$0.113  &121  &0.339  \\
		LSC2238  &  107.895$\pm$0.171  &  -27.265$\pm$0.169  &  0.158$\pm$0.099  &  -0.632$\pm$0.239  &  1.977$\pm$0.086  & 545  &0.383  \\
		LSC3029  &  124.006$\pm$0.169  &  -29.385$\pm$0.164  &  0.117$\pm$0.053  &  -1.853$\pm$0.261  &  2.082$\pm$0.270  & 568  &0.385  \\
		LSC4727  &  130.919$\pm$0.181  &  -29.677$\pm$0.169  &  0.149$\pm$0.082  &  -2.546$\pm$0.148  &  2.513$\pm$0.287  & 157  &0.369  \\
		LSC5588  &  92.743$\pm$0.181  &  37.855$\pm$0.180  &  0.256$\pm$0.141  &  0.822$\pm$0.106  &  -1.562$\pm$0.282  & 177  &0.374  \\
		LSC5969  &  113.190$\pm$0.158  &  -29.680$\pm$0.162  &  0.085$\pm$0.042  &  -1.108$\pm$0.229  &  2.241$\pm$0.285  &762  &0.393  \\
		LSC6714  &  108.764$\pm$0.168  &  -33.307$\pm$0.169  &  0.202$\pm$0.115  &  -1.021$\pm$0.264  &  2.935$\pm$0.119  &318  &0.368  \\
		\hline                                   
	\end{tabular}
\end{table}

\section{CMD fitting method and process}
\subsection{Fitting method}
The CMD fitting code employed in this work is $Powerful$ $CMD$ \citep{2017RAA....17...71L}, a reliable program for cluster SP synthesis. It was frequently-used to study the fundamental parameters and peculiar CMD morphologies of star clusters in the Milky Way \citep{2018Ap&SS.363...97L, 2018NewA...64...61L, 2021ApJS..253...38L, 2022ApJS..259...19L, 2023ApJS..265....3L} and MCs \citep{2020Ap&SS.365..134L, 2023MNRAS.525..827L}. Owing to the application of advanced stellar population synthesis (ASPS) \citep{2012ApJ...761L..22L, 2015ApJ...802...44L, 2016ApJS..225....7L}, the highlight of the code is the consideration of age spread, binaries and rotating stars. The model applies binary fractions from 0 to 1 with alternative intervals and six rotating star fractions (0, 0.1, 0.3, 0.5, 0.7, 1.0). A rapid binary-evolution algorithm of \cite{2002MNRAS.329..897H} is adopted to calculate binary evolution, and the rotation rate distribution is taken from \cite{2007A&A...463..671R}. Eight values (0.0001, 0.0003, 0.001, 0.004, 0.008, 0.01, 0.02, 0.03) of metallicities and 151 ages (0 - 15\,Gyr) are contained in this model.

In order to obtain the best-fitting parameters of star clusters, a weight avarage difference (WAD) is used to estimate the goodness of fit, as the CMD is divided into numerous grids. The value of WAD can be calculated by 
\begin{equation}\label{eq1}
	WAD = \Sigma|f_{ob} - f_{th}|,
\end{equation}
the $f_{ob}$ and $f_{th}$ are star fractions of observation and theory in each grid. After testing various statistical methods, the fitting result determined by WAD is the best. Therefore, we adopt the best-fitting result according to the minimum WAD value. Note that the photometric system applied for $Powerful$ $CMD$ is the Johnson-Cousins UBVRI system.
\subsection{CMD fitting process}
Firstly, we transfer the magnitudes from $G$ and $G_{RP}$ bands to $V-I$ and $V$ bands to implement the CMD fitting. The conversion formulas are given by \cite{2021A&A...649A...3R}. To minimize the uncertainty from magnitude transformation, we choose the following equations:
\begin{equation}\label{eq2}
	\begin{split}
		G - V = &-0.01597 - 0.02809(V - I_C) - 0.2483(V - I_C)^{2} +\\
		&0.03656(V - I_C)^{3} - 0.002939(V - I_C)^{4},\\
		G_{RP} - V = &0.01868 - 0.9028(V - I_C) - 0.005321(V - I_C)^{2} -\\
		&0.004186(V - I_C)^{3},
	\end{split}
\end{equation}
because of the minimum values of $\sigma$ and the closest distance from the curve of photometric relationships. To avoid the cause of magnitude transformation in reproducing the eMSTO feature, we compare the CMDs in $G$, $G_{BP}$, $G_{RP}$, $V-I$ and $V$ bands (see Figure 1). As can be seen, the eMSTOs exhibit clearly in all the bands. 
\begin{figure*}
	\centering
	\includegraphics[width=6in]{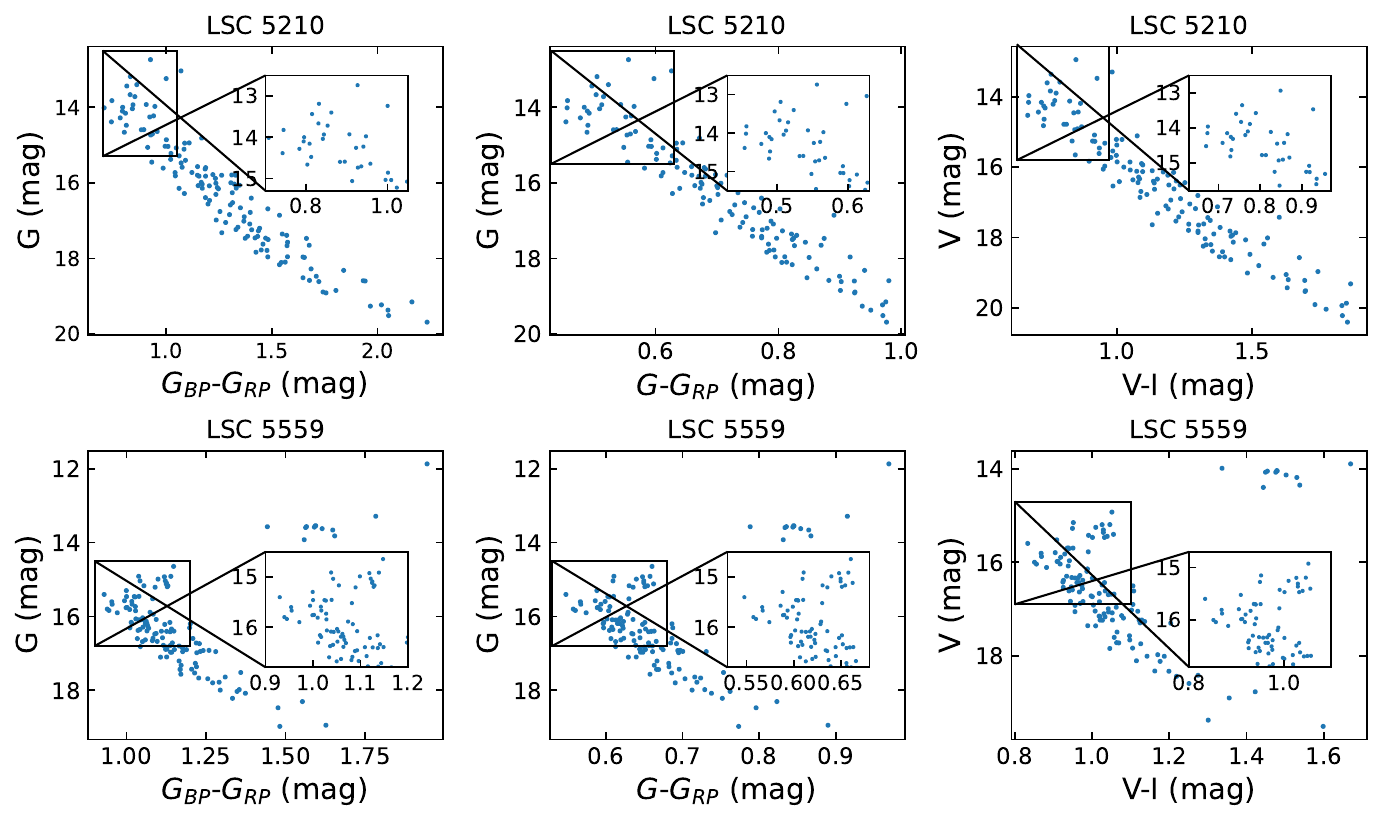}
	\caption{Observed CMDs of cluster samples in different bands as examples. The rectangles zoom in on eMSTO regions.}
\end{figure*}

Secondly, we divide the CMDs into 900 grids, including 30 colour bins and 30 magnitude bins. The number of stars in each grid is counted, and the grids with the largest star numbers in each magnitude bin (can be one or a few) are relatively dense regions. Due to the location (left edge or middle region of entire CMD) of best-fitting isochrone affects the accuracy of CMD fitting \citep{2023MNRAS.521.6284D}, we use these regions (as shown in Figure 2) for calculating the value of WAD. These regions are also helpful for determining the times of extended star formation. When other factors can not reproduce the eMSTO structure well, the eMSTO can be explained by multiple times of star formation within a small time interval or fewer times within a large time interval. We can limit the number of star formation times if the relatively dense regions at TO can be connected into obvious different curves.
\begin{figure*}
	\centering
	\includegraphics[width=5in]{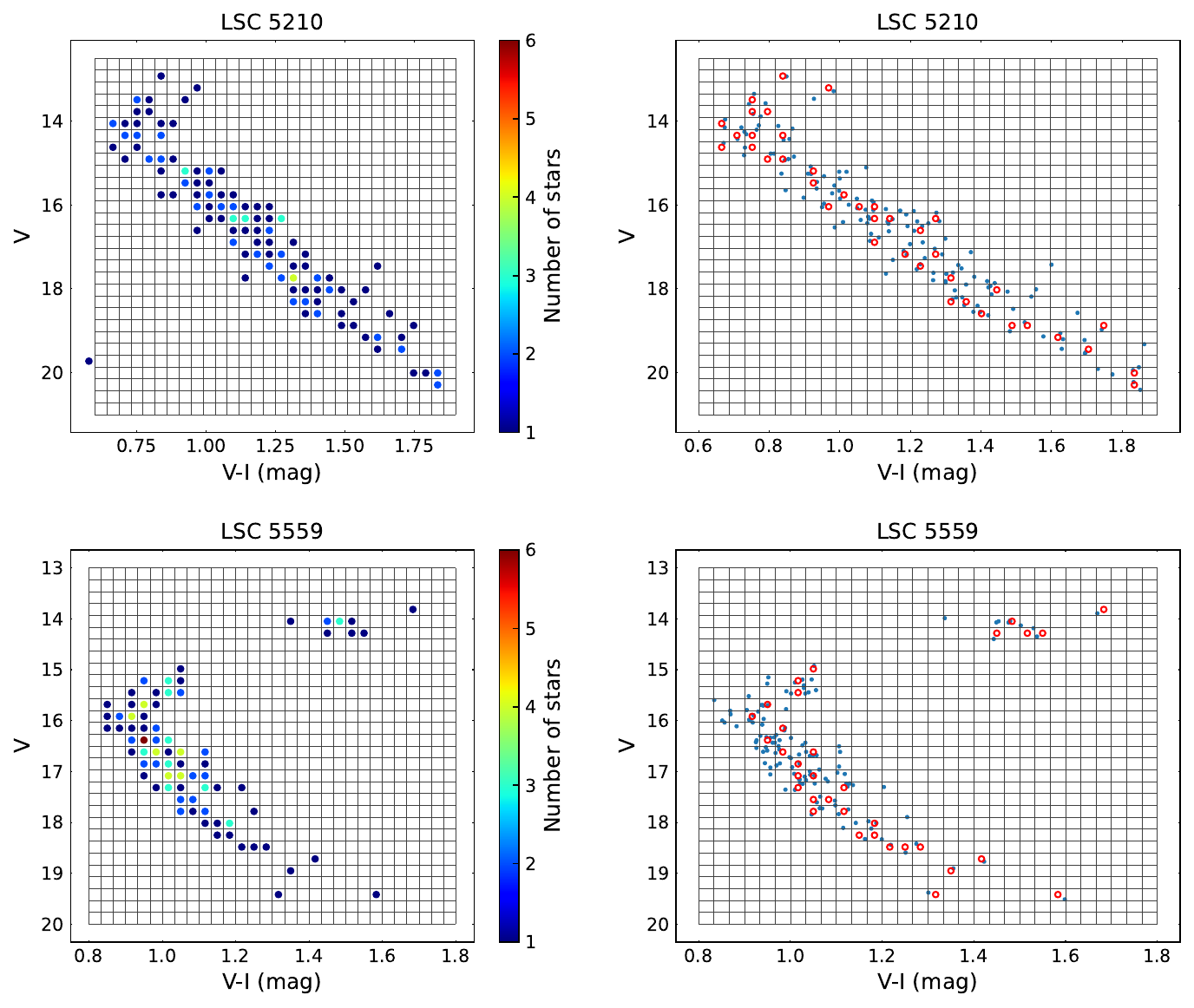}
	\caption{Examples of divided CMDs of cluster samples. Left panels show the number of stars in each grid. In right panels, blue dots are observed CMDs and red circles mark the relatively dense regions, which indicate the grids with the largest star numbers in each magnitude bin.}
\end{figure*}

Then, in contrast to the steps taken in our previous works, we fit the observed CMD using only single stars and check the theoretical CMD by eye to obtain the parameter range near the best-fitting result of each cluster. Finally, considering the effect of binaries, rotating stars and age spread, we classify seven different SP types to fit the observed CMDs by adjusting the parameters in the acquired ranges. The SP types respectively comprise only binaries ($b$), only rotating stars ($r$), only age spread ($a$), binary and rotating stars ($br$), binaries and age spread ($ba$), age spread and rotating stars ($ar$), and binary and rotating stars with age spread ($bar$). The minimum value of WAD corresponds to the best-fitting SP type that is mainly responsible for the eMSTO phenomenon.

\section{Results}
The cluster samples are classified into three types based on whether the observed CMDs have been entirely reproduced:

1. Type A: both main sequences and eMSTOs can be well-reproduced;

2. Type B: eMSTOs can be well-reproduced, main sequences cannot be well-reproduced;

3. Type C: other cases except B and C.

Within 11 OCs are Type A, six  OCs are Type B, and the rest are Type C. This work focuses on the fitting results of Type A samples. The fitted models of Types B and C individuals with relatively minor WAD values are displayed. We list the fitted SP type and the corresponding WAD value of each OC in Tables 2 and 3. As can be seen, all the cluster samples contain binaries. The intermediate-age OCs ($\geq$1.9\,Gyr) comprise no rotating stars since rotation almost does not affect stellar colour and magnitude in this stage \citep{2016ApJS..225....7L}. 

\begin{table}
	\caption{The fitted SP types and corresponding WAD values of Type A OCs. Type $b$, $a$ and $r$ denotes binaries, age pread and rotating stars.}       
	\label{table:2}      
	\centering                          
	\begin{tabular}{c c c c c c c c c c c c c}        
			\hline                
			ID & WAD & SP type &  Type &&&& ID & WAD & SP type &  Type\\
			\hline 
            LSC0667 & 26.6667 & bar & A &&&& LSC5048 & 27.027 & br & A \\
            LSC0667 & 28.7278 & ba & A &&&& LSC5048 & 27.456 & b & A \\
            LSC0667 & 30.4878 & br & A &&&& LSC5048 & 35.6757 & ar & A \\
            LSC0667 & 33.3333 & b & A &&&& LSC5048 & 36.3825 & r & A \\
            LSC0667 & 40.8163 & r & A &&&& LSC5048 & 36.3825 & a & A \\
            LSC0667 & 43.8312 & ar & A &&&& LSC5120 & 8.9556 & br & A \\
            LSC0667 & 45.4545 & a & A &&&& LSC5120 & 9.1577 & bar & A \\
            LSC3348 & 19.9523 & br & A &&&& LSC5120 & 9.3828 & ba & A \\
            LSC3348 & 21.3097 & b & A &&&& LSC5120 & 9.801 & b & A \\
            LSC3348 & 22.121 & bar & A &&&& LSC5120 & 10.1823 & ar & A \\
            LSC3348 & 22.9111 & ba & A &&&& LSC5120 & 10.4929 & a & A \\
            LSC3348 & 27.747 & r & A &&&& LSC5120 & 10.8771 & r & A \\
            LSC3348 & 32.5309 & ar & A &&&& LSC5210 & 14.2459 & br & A \\
            LSC3348 & 32.5309 & a & A &&&& LSC5210 & 14.6793 & bar & A \\
            LSC3491 & 8.4492 & bar & A &&&& LSC5210 & 15.5583 & ba & A \\
            LSC3491 & 8.6804 & ba & A &&&& LSC5210 & 16.2340 & b & A \\
            LSC3491 & 8.8645 & br & A &&&& LSC5210 & 17.2469 & ar & A \\
            LSC3491 & 9.6219 & b & A &&&& LSC5210 & 18.9210 & r & A \\
            LSC3491 & 11.552 & ar & A &&&& LSC5210 & 19.0204 & a & A \\
            LSC3491 & 11.6402 & r & A &&&& LSC5301 & 7.3026 & bar & A \\
            LSC3491 & 11.9199 & a & A &&&& LSC5301 & 7.3329 & br & A \\
            LSC3803 & 8.9561 & br & A &&&& LSC5301 & 7.9212 & ba & A \\
            LSC3803 & 9.1425 & bar & A &&&& LSC5301 & 8.3592 & b & A \\
            LSC3803 & 9.9441 & ba & A &&&& LSC5301 & 9.3254 & r & A \\
            LSC3803 & 10.0987 & b & A &&&& LSC5301 & 9.385 & ar & A \\
            LSC3803 & 11.4411 & r & A &&&& LSC5301 & 9.6899 & a & A \\
            LSC3803 & 11.6625 & ar & A &&&& LSC5559 & 17.6171 & ba & A \\
            LSC3803 & 12.3932 & a & A &&&& LSC5559 & 18.2648 & b & A \\
            LSC4611 & 8.4351 & bar & A &&&& LSC5559 & 22.8848 & a & A \\
            LSC4611 & 8.4739 & br & A &&&& LSC5867 & 15.9498 & bar & A \\
            LSC4611 & 8.5282 & ba & A &&&& LSC5867 & 16.4453 & ba & A \\
            LSC4611 & 9.2059 & b & A &&&& LSC5867 & 16.8303 & br & A \\
            LSC4611 & 10.9864 & ar & A &&&& LSC5867 & 17.6305 & b & A \\
            LSC4611 & 11.3214 & a & A &&&& LSC5867 & 19.589 & r & A \\
            LSC4611 & 11.4829 & r & A &&&& LSC5867 & 19.9693 & ar & A \\
            LSC5048 & 24.9288 & bar & A &&&& LSC5867 & 20.1887 & a & A \\
            LSC5048 & 26.6236 & ba & A &&&&  &  &  &  \\
			\hline                                   
	\end{tabular}
\end{table}

\begin{table}
	\caption{Similar to Table 2, but for Type B and C OCs.}             
	\label{table:3}      
	\centering                          
	\begin{tabular}{c c c c c c c c c c c c c}        
		\hline                
		ID & WAD & SP type &  Type &&&& ID & WAD & SP type &  Type\\
		\hline 
		LSC0398 & 6.1455 & br & B &&&& LSC0623 & 4.9936 & b & C \\
		LSC0398 & 6.2867 & bar & B &&&& LSC0623 & 5.8093 & ar & C \\
		LSC0398 & 6.7649 & ba & B &&&& LSC0623 & 5.8243 & r & C \\
		LSC0398 & 6.9696 & b & B &&&& LSC0623 & 6.0278 & a & C \\
		LSC0398 & 8.4573 & ar & B &&&& LSC2064 & 16.1329 & ba & C \\
		LSC0398 & 8.5333 & r & B &&&& LSC2064 & 16.5737 & b & C \\
		LSC0398 & 8.7419 & a & B &&&& LSC2064 & 21.8196 & a & C \\
		LSC0574 & 6.8678 & ba & B &&&& LSC2238 & 8.9375 & b & C \\
		LSC0574 & 7.1937 & b & B &&&& LSC3029 & 5.6927 & bar & C \\
		LSC0574 & 9.0479 & a & B &&&& LSC3029 & 5.7651 & br & C \\
		LSC1134 & 9.3165 & ba & B &&&& LSC3029 & 5.8916 & ba & C \\
		LSC1134 & 9.5233 & b & B &&&& LSC3029 & 6.2088 & b & C \\
		LSC1134 & 12.5082 & a & B &&&& LSC3029 & 7.0141 & r & C \\
		LSC1323 & 9.1125 & ba & B &&&& LSC3029 & 7.2422 & ar & C \\
		LSC1323 & 9.621 & b & B &&&& LSC3029 & 7.4464 & a & C \\
		LSC4612 & 10.7741 & b & B &&&& LSC4727 & 15.7325 & b & C \\
		LSC5576 & 13.7371 & bar & B &&&& LSC5588 & 11.7349 & ba & C \\
		LSC5576 & 14.0915 & ba & B &&&& LSC5588 & 12.4103 & b & C \\
		LSC5576 & 14.7439 & br & B &&&& LSC5588 & 15.9416 & a & C \\
		LSC5576 & 15.4965 & b & B &&&& LSC5969 & 5.1336 & bar & C \\
		LSC5576 & 16.4282 & ar & B &&&& LSC5969 & 5.1729 & br & C \\
		LSC5576 & 16.7296 & a & B &&&& LSC5969 & 5.1804 & ba & C \\
		LSC5576 & 17.9888 & r & B &&&& LSC5969 & 5.5541 & b & C \\
		LSC0607 & 6.7575 & ba & C &&&& LSC5969 & 6.0983 & r & C \\
		LSC0607 & 7.0699 & b & C &&&& LSC5969 & 6.137 & ar & C \\
		LSC0607 & 8.6217 & a & C &&&& LSC5969 & 6.1582 & a & C \\
		LSC0623 & 4.4586 & bar & C &&&& LSC6714 & 8.6554 & b & C \\
		LSC0623 & 4.572 & br & C &&&& LSC6714 & 8.8555 & ba & C \\
		LSC0623 & 4.6726 & ba & C &&&&  &  &  &  \\
		\hline                                   
	\end{tabular}
\end{table}

However, WAD is not the unique criterion for evaluating fitting, as any statistical indicator may have shortcomings. We revisit the comparison of observed and theoretical CMDs of Type A samples by visual check because most best-fitting SP types include age spread. As shown in Figure 3, the eMSTO phenomena of OCs LSC0667 and LSC 5048 can be reproduced well by both $ba$ and $bar$ SP types, but the $ba$ SP type performs a better fitting at brighter magnitudes. In the middle panels, the $br$ SP type cannot reproduce the data at all. It suggests that the effect of age spread is much greater than that of stellar rotation on eMSTO phenomena. Thus the best-fitting SP types of LSC0667 and LSC5048 tend to be $ba$ type. In Figure 4, OCs LSC3491 and 5301 can be explained by both $br$ and $ba$ SP types, although the $bar$ SP type appears a better fittting to the observed data. We cannot rule out any possibility, as the three factors could be the reasons for eMSTOs. For LSC5210, the $bar$ SP type reproduces the morphology of eMSTO very well. As a result, we prefer to use $bar$ SP type for interpreting the eMSTOs of LSC3491, 5210 and 5301. Meanwhile, the combination of binaries and rotating stars is mainly responsible for the eMSTO features of LSC4611 and 5867, similar to the result in \cite{2012ApJ...761L..22L}. The effect of age spread cannot be ignored either. 

In addition, the result of LSC5559 seems weird that stellar rotation is useless in reproducing eMSTO morphology, although the age ($\sim$1.6\,Gyr) is younger than the studied ranges of previous works. Since only a small fraction of stars have stellar rotation rates ($\omega$ = $\Omega$/$\Omega$$_{crit}$) larger than 0.7 in \cite{2007A&A...463..671R}, suggesting that $Powerfule$ $CMD$ code is difficult to fit fast rotators. We therefore apply another widely-used stellar evolutionary isochrones, PARSEC V2.0\footnote{\url{http://stev.oapd.inaf.it/cgi-bin/cmd_3.7}} \citep{2012MNRAS.427..127B, 2022A&A...665A.126N}, for SP studying. The method of isochrone fitting is explained in \cite{2023ApJS..267...34H}. As provided in Figure 5, the eMSTO region can be well-fitted by isochrones with $\omega$ from 0 to 0.99. The dense grids (with relatively large star fractions) are also well-reproduced by isochrone with $\omega$$_i$ = 0.80. It suggests that stellar rotation is the main reason for the eMSTO feature of LSC5559. The obvious colour dispersion along the main sequence below the turn-off region is mainly caused by binaries.
The best-fitting SP types of the other three OCs of Type A are plotted in Figure 6. We can see that binaries and rotating stars can explain the eMSTOs of LSC3348, 3803 and 5120. 

Table 4 concludes the fitted parameters and SP types of all the OCs. 
According to the fitting results of Type A samples, we find that binaries and star formation can qualitatively match eMSTO structures of young Galactic OCs ($\sim$0.0 and 0.6\,Gyr). The observed eMSTOs of OCs within an older age range can be explained by stellar binarity and rotation. It seems necessary to consider the combination of binaries, rotating stars, and age spread because it can fit three OCs better. In addition, eight OCs in Type A appear to contain a rotating star fraction larger than 0.5. It implicates that most member stars in MSTO regions of OCs in this work are slow rotators, apart from those of LSC5559. 
The main sequences and MSTOs in observed CMDs of Type B and C are extremely broad, possibly resulting in errors in observation and data processing. As age spread reproduces all the eMSTOs, the fitting results of the two types are not very accurate.

\begin{figure*}
	\centering
	\includegraphics[width=5in]{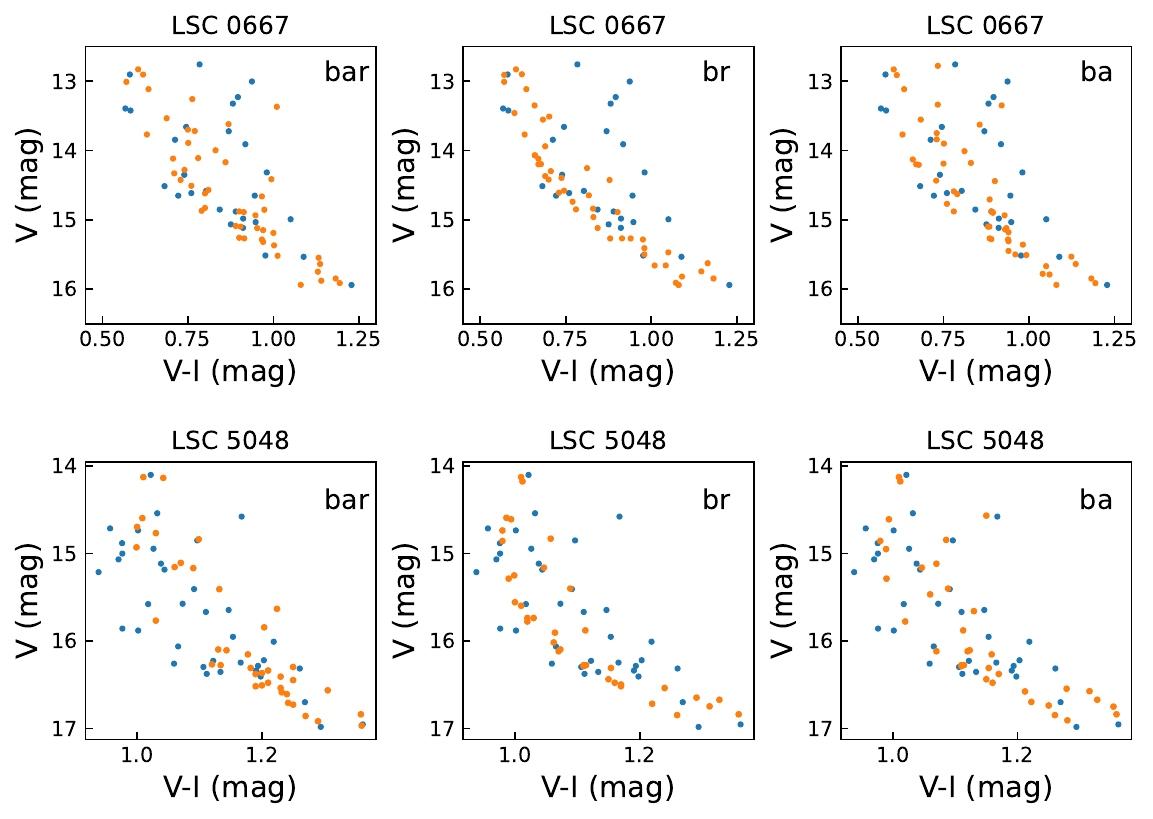}
	\caption{The comparisons of observed and best-fitting eMSTO structures in CMDs of two OCs in Type A. Blue dots indicate observed data and orange dots are for theoretical data. Panels in each row represent different SP types from left to right. Type $b$, $a$ and $r$ indicate binaries, age pread and rotating stars respectively.}
\end{figure*}
\begin{figure*}
	\centering
	\includegraphics[width=5.5in]{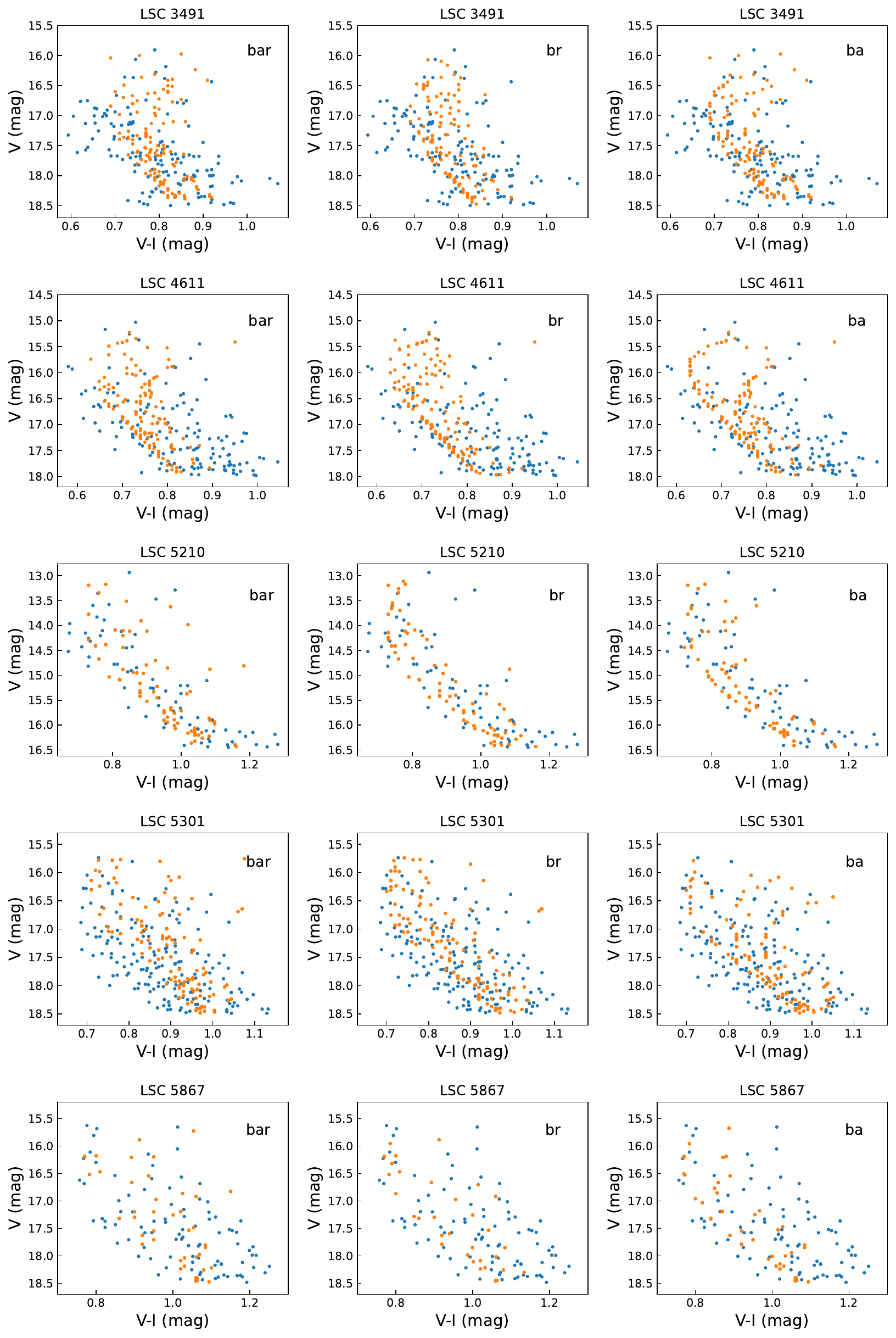}
	\caption{Similar to Figure3, but for other five OCs.}
\end{figure*}
\begin{figure*}
	\centering
	\includegraphics[width=2.8in]{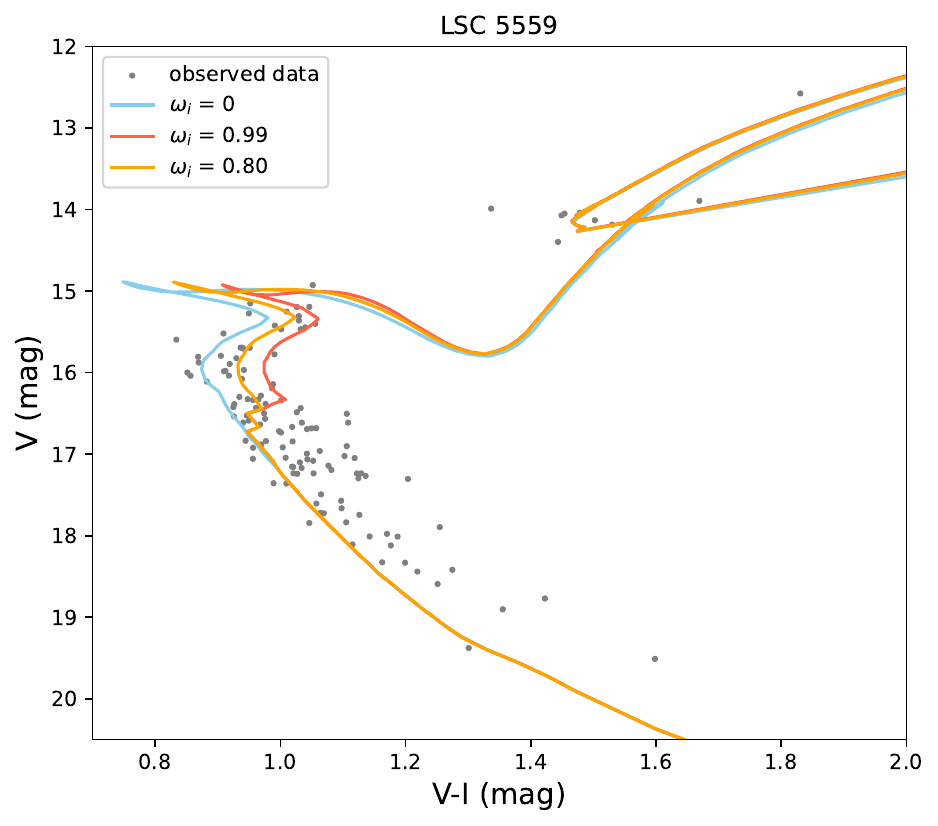}
	\caption{Comparison of observed CMD and fitted isochrones of LSC5559 from PARSEC V2.0.}
\end{figure*}
\begin{figure*}
	\centering
	\includegraphics[width=5.5in]{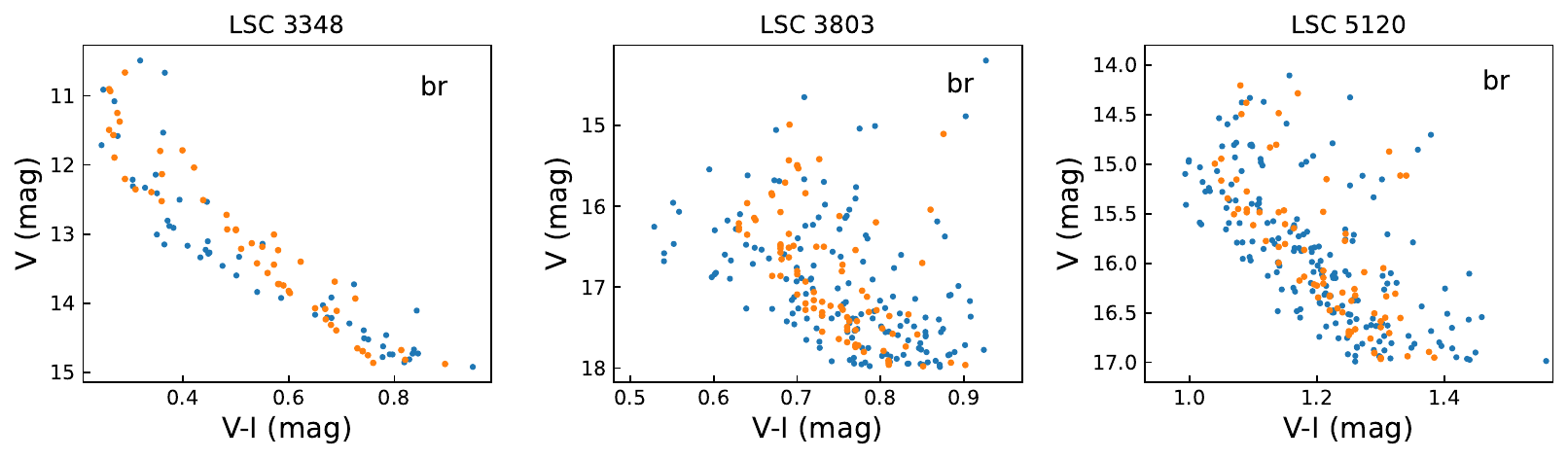}
	\caption{Comparisons of observed and best-fitting eMSTO structures in CMDs of LSC3348, 3803 and 5120. Blue and orange dots represent observed and theoretical data respectively.}
\end{figure*}
\begin{table*}
	\caption{The best-fitting parameters and SP types of all the OCs in this work. $Z$, $m-M$, $E(V-I)$, $f_b$, $f_r$ demonstrate metallicity, distance modulus, colour excess, binary fraction and rotating star fraction respectively. Note that $Age$ indicates the age of the youngest population in the OC. $NUM_{SF}$ indicates the times of star formation and $Age_s$ is age difference between two adjacent star formations.}             
	\label{table:4}      
	\centering 
	\scalebox{0.85}{                         
	\begin{tabular}{c c c c c c c c c c c c c}        
		\hline                 
		ID  &  $Z$  &  $m-M$  &  $E(V-I)$  &  $f_b$  &  $NUM_{SF}$  &  $Age$  &  $f_r$  &  WAD  &  SP type  &  $Age_s$  &  Type  &  Name  \\
		&  &  [mag]  &  [mag]  &  &  &  [Gyr]  &  &  &  &  [Gyr]  &  &  \\
		\hline                
		LSC0667  &  0.0040   &  10.90   &  0.75   &  0.54   &  3  &  0.0   &  0.0   &  27.2222  &  ba  &  0.5   &  A  &  Teutsch226, UBC497  \\
		LSC3348  &  0.0080   &  9.90   &  0.21   &  0.52   &  1  &  0.6   &  0.7   &  19.9523  &  br  &  0.0   &  A  &  Collinder208, Melotte97  \\
		LSC3491  &  0.0100   &  13.90   &  0.36   &  0.47   &  3  &  1.5   &  0.5   &  8.4492  &  bar  &  0.2   &  A  &  -  \\
		LSC3803  &  0.0300   &  14.00   &  0.36   &  0.55   &  1  &  0.8   &  1.0   &  8.9561  &  br  &  0.0   &  A  &  -  \\
		LSC4611  &  0.0100   &  13.00   &  0.28   &  0.55   &  1  &  1.6   &  0.7   &  8.4739  &  br  &  0.0   &  A  &  -  \\
		LSC5048  &  0.0040   &  11.60   &  0.98   &  0.52   &  2  &  0.6   &  0.0   &  26.6236  &  ba  &  0.4   &  A  &  UBC299  \\
		LSC5120  &  0.0200   &  11.86   &  0.89   &  0.54   &  1  &  0.7   &  0.7   &  8.9556  &  br  &  0.0   &  A  &  MWSC2261, Ruprecht112  \\
		LSC5210  &  0.0300   &  12.06   &  0.68   &  0.51   &  2  &  0.4   &  0.7   &  14.6793  &  bar  &  0.2   &  A  &  FoF1994, UBC288  \\
		LSC5301  &  0.0300   &  14.20   &  0.56   &  0.53   &  2  &  0.6   &  1.0   &  7.3026  &  bar  &  0.2   &  A  &  -  \\
		LSC5559  &  0.0140   &  13.40   &  0.43   &  0.00   &  1  &  1.6   &  -  &  -  &  r  &  0.0   &  A  &  vdBergh49  \\
		LSC5867  &  0.0200   &  14.00   &  0.70   &  0.54   &  1  &  0.5   &  0.7   &  16.8303  &  br  &  0.0   &  A  &  -  \\
		LSC0398  &  0.0200   &  13.45   &  0.51   &  0.55   &  1  &  0.9   &  1.0   &  6.1455  &  br  &  0.0   &  B  &  -  \\
		LSC0574  &  0.0200   &  12.80   &  0.17   &  0.55   &  2  &  1.9   &  0.0   &  6.8678  &  ba  &  1.2   &  B  &  -  \\
		LSC1134  &  0.0200   &  13.25   &  0.13   &  0.51   &  2  &  2.1   &  0.0   &  9.3165  &  ba  &  0.6   &  B  &  -  \\
		LSC1323  &  0.0200   &  13.92   &  0.24   &  0.54   &  2  &  1.4   &  0.0   &  9.1125  &  ba  &  0.5   &  B  &  -  \\
		LSC4612  &  0.0100   &  13.30   &  0.15   &  0.53   &  1  &  2.5   &  0.0   &  10.7741  &  b  &  0.0   &  B  &  -  \\
		LSC5576  &  0.0200   &  13.25   &  0.36   &  0.52   &  2  &  0.9   &  0.7   &  13.7371  &  bar  &  0.2   &  B  &  -  \\
		LSC0607  &  0.0200   &  13.95   &  0.14   &  0.54   &  2  &  1.9   &  0.0   &  6.7575  &  ba  &  0.5   &  C  &  -  \\
		LSC0623  &  0.0100   &  14.93   &  0.32   &  0.55   &  2  &  1.6   &  1.0   &  4.4586  &  bar  &  0.5   &  C  &  -  \\
		LSC2064  &  0.0200   &  14.10   &  0.20   &  0.49   &  2  &  1.6   &  0.0   &  16.1329  &  ba  &  0.3   &  C  &  -  \\
		LSC2238  &  0.0200   &  13.80   &  0.15   &  0.54   &  1  &  1.9   &  0.0   &  8.9375  &  b  &  0.0   &  C  &  -  \\
		LSC3029  &  0.0080   &  13.80   &  0.48   &  0.53   &  2  &  1.5   &  0.7   &  5.6927  &  bar  &  0.5   &  C  &  -  \\
		LSC4727  &  0.0080   &  13.80   &  0.30   &  0.54   &  1  &  2.1   &  0.0   &  15.7325  &  b  &  0.0   &  C  &  -  \\
		LSC5588  &  0.0100   &  13.40   &  0.34   &  0.54   &  2  &  1.9   &  0.0   &  11.7349  &  ba  &  0.4   &  C  &  -  \\
		LSC5969  &  0.0100   &  14.33   &  0.29   &  0.51   &  3  &  1.5   &  0.7   &  5.1336  &  bar  &  0.4   &  C  &  -  \\
		LSC6714  &  0.0100   &  14.15   &  0.19   &  0.54   &  1  &  1.8   &  0.0   &  8.6554  &  b  &  0.0   &  C  &  -  \\
		\hline                                   
	\end{tabular}}
\end{table*}

\section{Conclusion}
This paper declares 26 Galactic OCs with unstudied eMSTOs and explores the reasons for their peculiar structures. We select twenty new and six known OCs as cluster samples from the LSC catalog of our previous work \citep{li2024bsec} and divide them into three types, i.e., Type A, Type B, and Type C. Stellar binarity, rotation, and age spread are taken into account for studying the leading causes for eMSTO features. The isochrones of $Powerful$ $CMD$ and PARSEC V2.0 are used to obtain the best-fitting SP types and fundamental parameters. The main results are summarized as follows:
\begin{itemize}
	\item[$\bullet$]The results of Type A (which includes 11 OCs) are reliable, as all the CMD morphologies of Type B and C are not well-reproduced.
	\item[$\bullet$]Binaries play a significant role in explaining the eMSTO structures.
	\item[$\bullet$]Age spread and stellar binarity can recover the eMSTOs of young OCs ($\leq$0.6\,Gyr).
	\item[$\bullet$]The effects of stellar rotation and binaries are mainly responsible for eMSTOs for most OCs in Type A, but the impact of age spread cannot be excluded because of the better fitting to observed CMDs.
	\item[$\bullet$]Most member stars in eMSTO regions of Type A OCs are slow-rotating stars.
\end{itemize}
With the search, identification and redetermination of more OCs in the Milky Way, an increasing number of Galactic OCs are found to harbor eMSTO structures. In future work, we will continue to explore the structure and provide more accurate research on the SP synthesis of Galactic star clusters.

\normalem
\begin{acknowledgements}
We thank Dr. Zhihong He and Mr. Xiaochen Liu for helping perform the PARSEC isochrone fitting. This work has been supported by Yunnan Academician Workstation of Wang Jingxiu (202005AF150025), China Manned Space Project (NO.CMS-CSST-2021-A08), GH project (ghfund202302019167), and the Natural Science Foundation of Yunnan Province (No. 202201BC070003).

This work has made use of data from the European Space Agency (ESA) mission
{\it Gaia} (\url{https://www.cosmos.esa.int/gaia}), processed by the {\it Gaia}
Data Processing and Analysis Consortium (DPAC,
\url{https://www.cosmos.esa.int/web/gaia/dpac/consortium}). Funding for the DPAC
has been provided by national institutions, in particular the institutions
participating in the {\it Gaia} Multilateral Agreement.

\end{acknowledgements}
  
\bibliographystyle{raa}
\bibliography{bibtex}

\end{document}